\documentclass[a4paper]{PoS}

\usepackage{amsmath,color,amscd,subfigure}
\usepackage{graphicx}
\usepackage[english]{babel}
\usepackage{listings}
\usepackage{cite}

\title{AVX-512 extension to OpenQCD 1.6}
\ShortTitle{AVX-512 extension to OpenQCD 1.6}

\author{\speaker{Jarno Rantaharju}\\
        Swansea Academy of Advanced Computing, Swansea University, SA1 8EP, UK\\
        E-mail: \email{j.m.o.rantaharju@swansea.ac.uk}
}
        
\author{{Ed Bennett}\\
        Swansea Academy of Advanced Computing, Swansea University, SA1 8EP, UK\\
        E-mail: \email{e.j.bennett@swansea.ac.uk}
}
        
\author{{Mark Dawson}\\
        Swansea Academy of Advanced Computing, Swansea University, SA1 8EP, UK\\
        E-mail: \email{mark.dawson@swansea.ac.uk}
}
        
\author{{Michele Mesiti}\\
        Swansea Academy of Advanced Computing, Swansea University, SA1 8EP, UK\\
        E-mail: \email{michele.mesiti@swansea.ac.uk}
}        

\abstract{
We publish an extension of openQCD-1.6 with AVX-512 vector instructions using Intel intrinsics.
Recent Intel processors support extended instruction sets with operations on 512-bit wide vectors, increasing both the capacity for floating point operations and register memory.
Optimal use of the new capabilities requires reorganising data and floating point operations into these wider vector units.
We report on the implementation and performance of the AVX-512 OpenQCD extension on clusters using Intel Knights Landing and Xeon Scalable (Skylake) CPUs. In complete HMC trajectories with physically relevant parameters we observe a performance increase of 5\% to 10\%.
}

\FullConference{The 36th Annual International Symposium on Lattice Field Theory - LATTICE2018\\
                22-28 July, 2018\\
                Michigan State University, East Lansing, Michigan, USA.}

\begin{document}

\section{Introduction}

The newest generations of Intel processors, Xeon Scalable (Skylake) and Xeon Phi (Knights Landing), extend the current standard of AVX2 vector instructions with the 512-bit wide AVX-512 instruction sets \cite{AVX512}. The width of registers is similarly increased and, in addition, the number of floating point registers is doubled to 32. Compilers can make use of the new instructions, but targeted code is required to reach optimal performance. Efficient vectorisation can lead to a doubling of the available register memory and floating point capability

OpenQCD-1.6 \cite{openqcd} already includes optional AVX2 and SSE targeted implementations and an extension targeting BlueGene/Q also exists \cite{BGQ}.
Following the logic of these extensions we have reimplemented several performance critical functions using Intel intrinsic instructions for AVX-512 vectors. These are mainly the Dirac operator and the vector operations necessary for the conjugate gradient algorithm.

Here we publish the extension to openQCD-1.6 \cite{sa2c-github-page}. In addition, we report scaling studies performed for OpenQCD-FASTSUM, the FASTSUM collaboration's extension of openQCD \cite{FASTSUM}, with the AVX-512 implementation.

\section{Implementation}

The implementation of the extension is guided by the expectation that lattice QCD simulations are memory bandwidth bound. The performance of the application is limited by the memory bandwidth between the processor and different cache levels rather than the capacity for floating point operations.
The same assumption guides the existing AVX2, SSE and BlueGene/Q targeted implementations. Gauge matrices and spinors of the Wilson formulation are stored in memory as structures and SIMD vectors are constructed out of spinor degrees of freedom.
Our implementation combines spinor and direction indices to construct the 512-bit vectors required.

The construction of a vector does not require any rearrangement of the data before loading into registers. However, different Dirac and directional indices are often handled differently, increasing the number of floating point instructions required. Since the primary objective is to optimise memory use, we consider this acceptable.

In the SSE and AVX2 extensions, direct insertions of assembly code are used to achieve full control over the compiled code. We use Intel intrinsic instructions. Compilers generally replace these routines with assembly instructions in a one-to-one correspondence. Intrinsic functions offer more freedom in choosing the optimal compiler and allows easier porting to different processor types. They leave the compiler with the task of choosing the optimal instruction for each operation and assigning data to registers. This is especially important since the number of registers is increased in Xeon Scalable CPUs \cite{AVX512}. Using prewritten assembly code would confine the extension only to future processors with the higher register count.

We implement several core functions, including the Dirac operator, the application of the Sheikholeslami-Wohlert term, and several linear algebra functions. The extension is activated using the \lstinline{AVX512} preprocessor flag, similarly to the existing \lstinline{AVX2} and \lstinline{SSE} preprocessor flags. When the flags are combined, the AVX-512 implementation is used when available.

\section{Benchmarking}

\begin{table} \centering
\begin{tabular}{|c|c|c|c|c|c|c|c|}
\hline
Volume  &  \multicolumn{3}{|c|}{ Knights Landing } & \multicolumn{3}{|c|}{ Skylake } \\
 & AVX-512 & AVX2 & Speedup & AVX-512 & AVX2 & Speedup \\
\hline
\multicolumn{7}{|l|}{ Single Precision } \\
\hline
4 $\times$4$\times$4$\times$4 & 8521 & 5455 & 1.56 & 36177 &  29839 & 1.21 \\
8 $\times$4$\times$4$\times$4 & 6276 & 4130 & 1.52 & 34649 &  27769 & 1.25 \\
8 $\times$8$\times$4$\times$4 & 6063 & 4042 & 1.50 & 36167 &  29289 & 1.23 \\
8 $\times$8$\times$8$\times$4 & 5286 & 3791 & 1.39 & 34894 &  28476 & 1.23 \\
8 $\times$8$\times$8$\times$8 & 5088 & 3721 & 1.37 & 26408 &  21617 & 1.22 \\
16$\times$8$\times$8$\times$8 & 4506 & 3338 & 1.35 & 25300 &  19180 & 1.32 \\
\hline
\multicolumn{7}{|l|}{ Double precision } \\
\hline
4 $\times$4$\times$4$\times$4 & 6164 & 3725 & 1.65 & 26737 & 24681 & 1.08 \\
8 $\times$4$\times$4$\times$4 & 4105 & 2857 & 1.44 & 26690 & 24609 & 1.08 \\
8 $\times$8$\times$4$\times$4 & 3533 & 2517 & 1.40 & 26521 & 19687 & 1.35 \\
8 $\times$8$\times$8$\times$4 & 3296 & 2421 & 1.36 & 25267 & 19312 & 1.31 \\
8 $\times$8$\times$8$\times$8 & 3191 & 2405 & 1.33 & 18772 & 14471 & 1.29 \\
16$\times$8$\times$8$\times$8 & 2911 & 2131 & 1.37 & 15513 & 15125 & 1.03 \\
\hline
\end{tabular}
\caption{ \label{singlecoretable} The performance of the functions Dw and Dw\_dble (single and double precision respectively) in Mflops on single Knights Landing and Skylake cores. }
\end{table}

\begin{figure}
\centering
\includegraphics[width=0.49\linewidth]{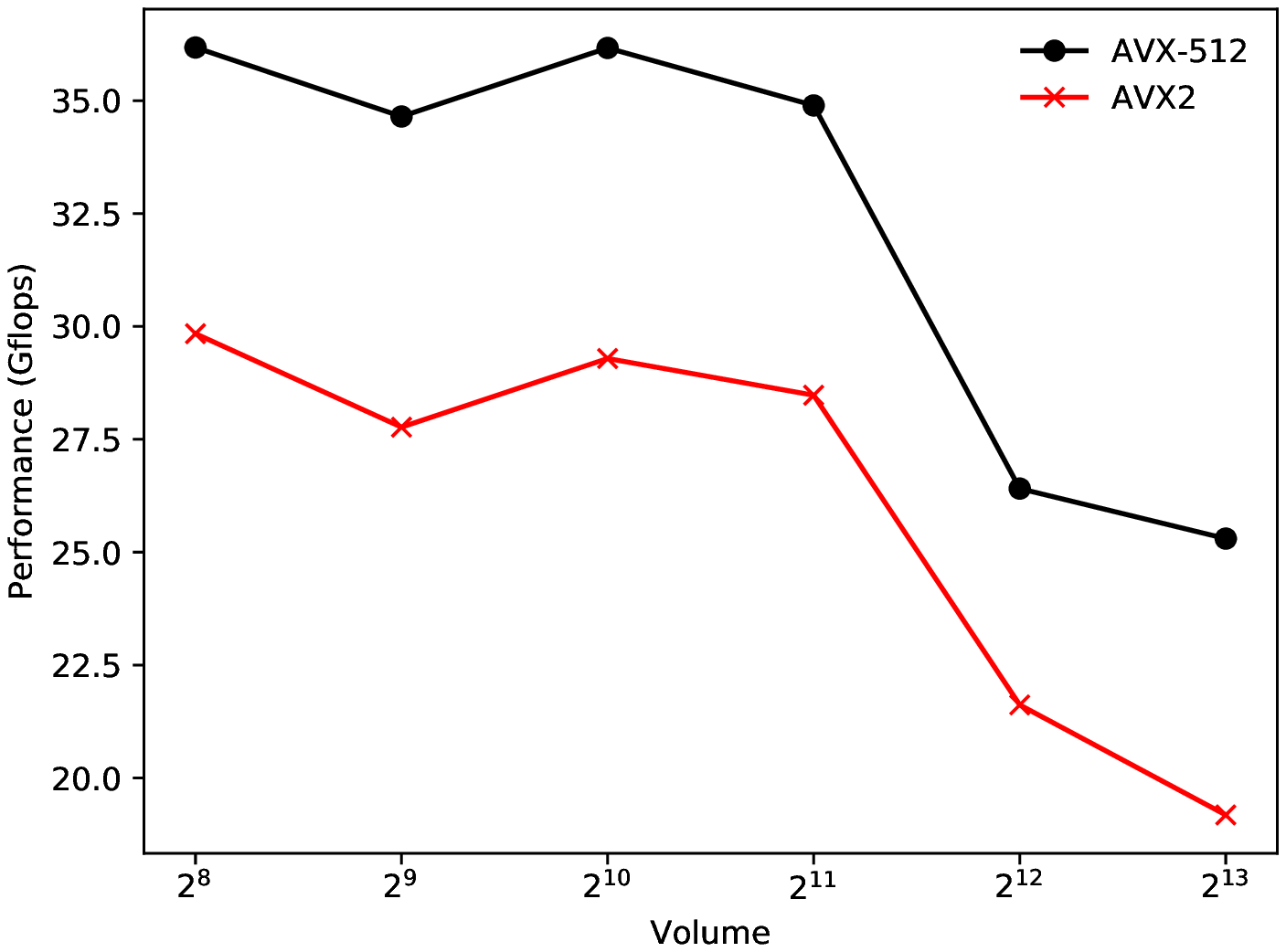}
\includegraphics[width=0.49\linewidth]{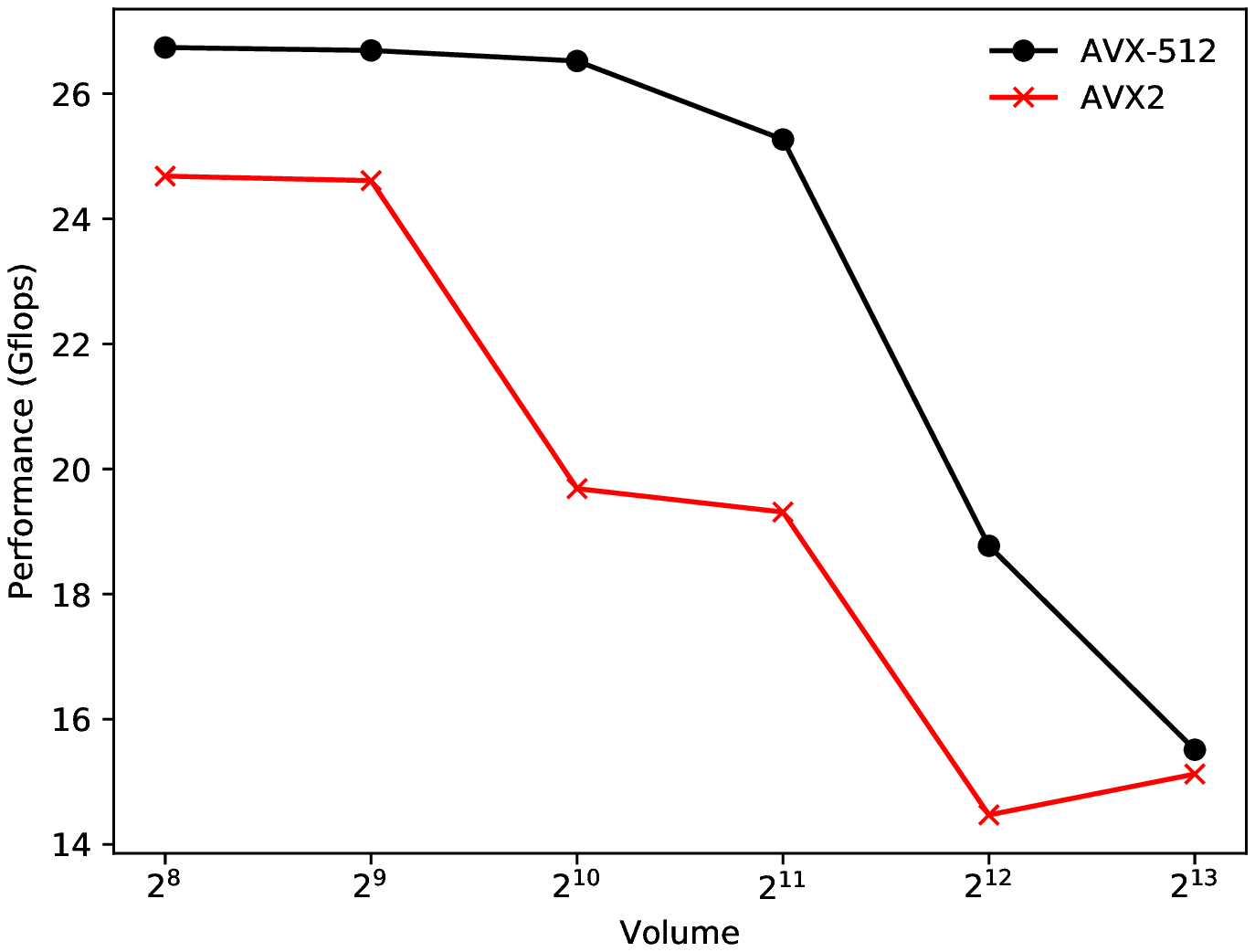}
\caption{ The performance of Dw() (left), Dw\_dble() (right) performance measures run on a single Skylake core using the AVX-512 and AVX2 implementations. }
\label{fig:singlecore}
\end{figure}

\begin{figure}
 \centering
 \includegraphics[width=0.49\linewidth]{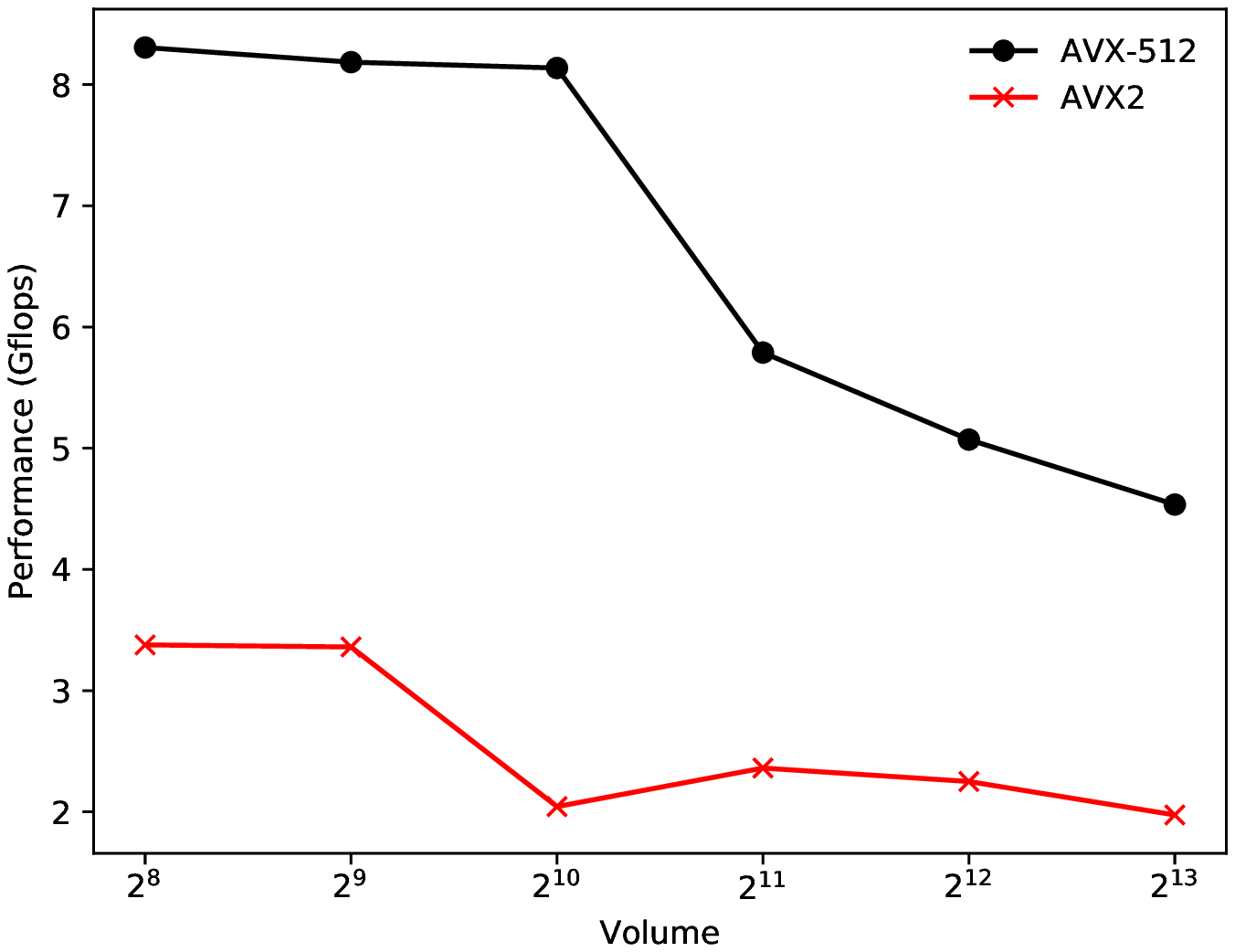}
 \includegraphics[width=0.49\linewidth]{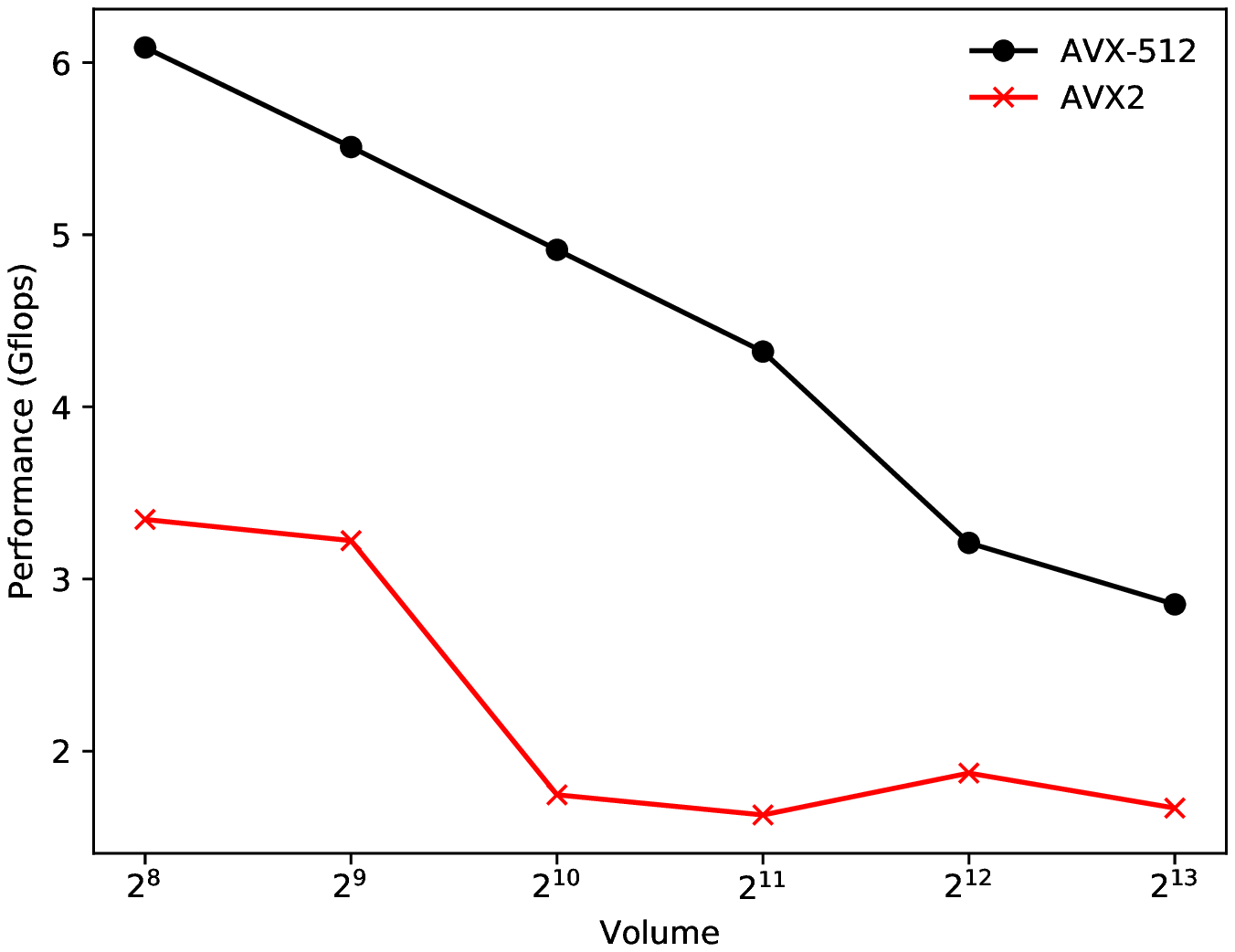}
 \caption{The Dw() (left), Dw\_dble() (right) performance measures on a single KNL core using the AVX-512 and AVX2 implementations. }
 \label{fig:singlecore_phi}
\end{figure}

We run several performance tests on the FASTSUM extension of OpenQCD 1.6 with and without the AVX-512 implementation on Cineca Marconi A2 cluster Intel Knights Landing nodes and the Supercomputing Wales Sunbird cluster with Intel Skylake nodes. On Sunbird we use the Intel C Compiler to build the AVX-512 implementation with the compiler flags
\begin{lstlisting}
-std=c89 -xCORE-AVX512 -mtune=skylake -O3 -DAVX512 -DAVX -DFMA3 -DPM.
\end{lstlisting}
The original AVX2 version is compiled with
\begin{lstlisting}
-std=c89 -xCORE-AVX512 -march=skylake -O3 -DPM -DAVX -DFMA3 -DPM.
\end{lstlisting}
On the Knights Landing cluster the AVX-512 version is compiled with
\begin{lstlisting}
-std=c89 -xMIC-AVX512 -O3 -DAVX512 -DAVX -DFMA3 -DPM
\end{lstlisting}
and compared against the AVX2 version compiled using
\begin{lstlisting}
-std=c89 -xMIC-AVX512 -O3 -DAVX -DFMA3 -DPM.
\end{lstlisting}

Firstly, we have measured the single core performance of the Dirac operator itself using the timing tools provided in the openQCD code. These tests do not account for memory dependencies, but only measure floating point performance. The performance in Mflops per second is shown in Tab. \ref{singlecoretable} and in Fig. \ref{fig:singlecore} and \ref{fig:singlecore_phi}. With small lattice sizes we observe a significant improvement, exceeding a factor of two with certain cases. Naturally the improvement is smaller in a realistic test case due to data dependencies and MPI communication.

\begin{figure}
 \centering
 \includegraphics[width=0.49\linewidth]{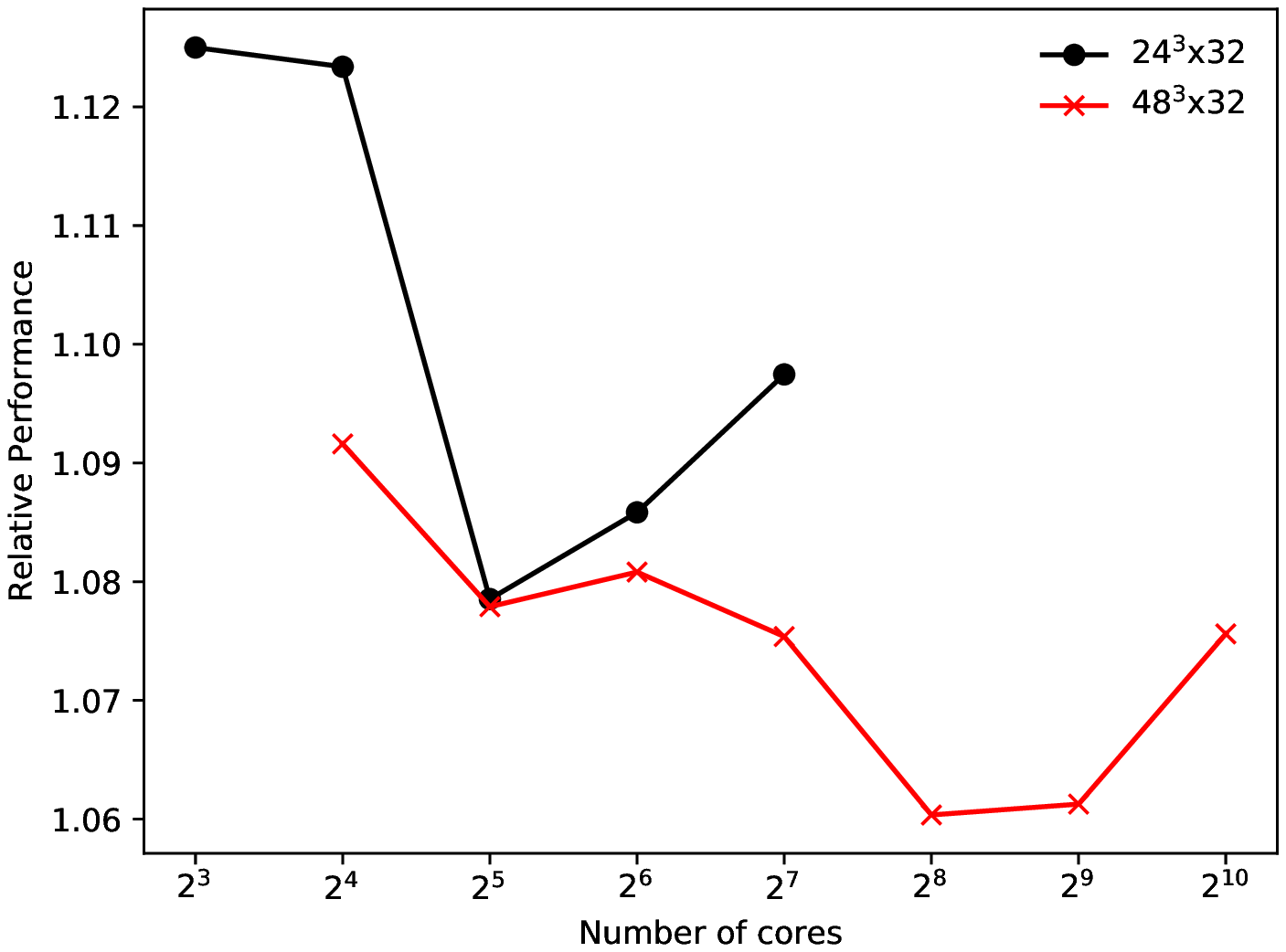}
 \includegraphics[width=0.49\linewidth]{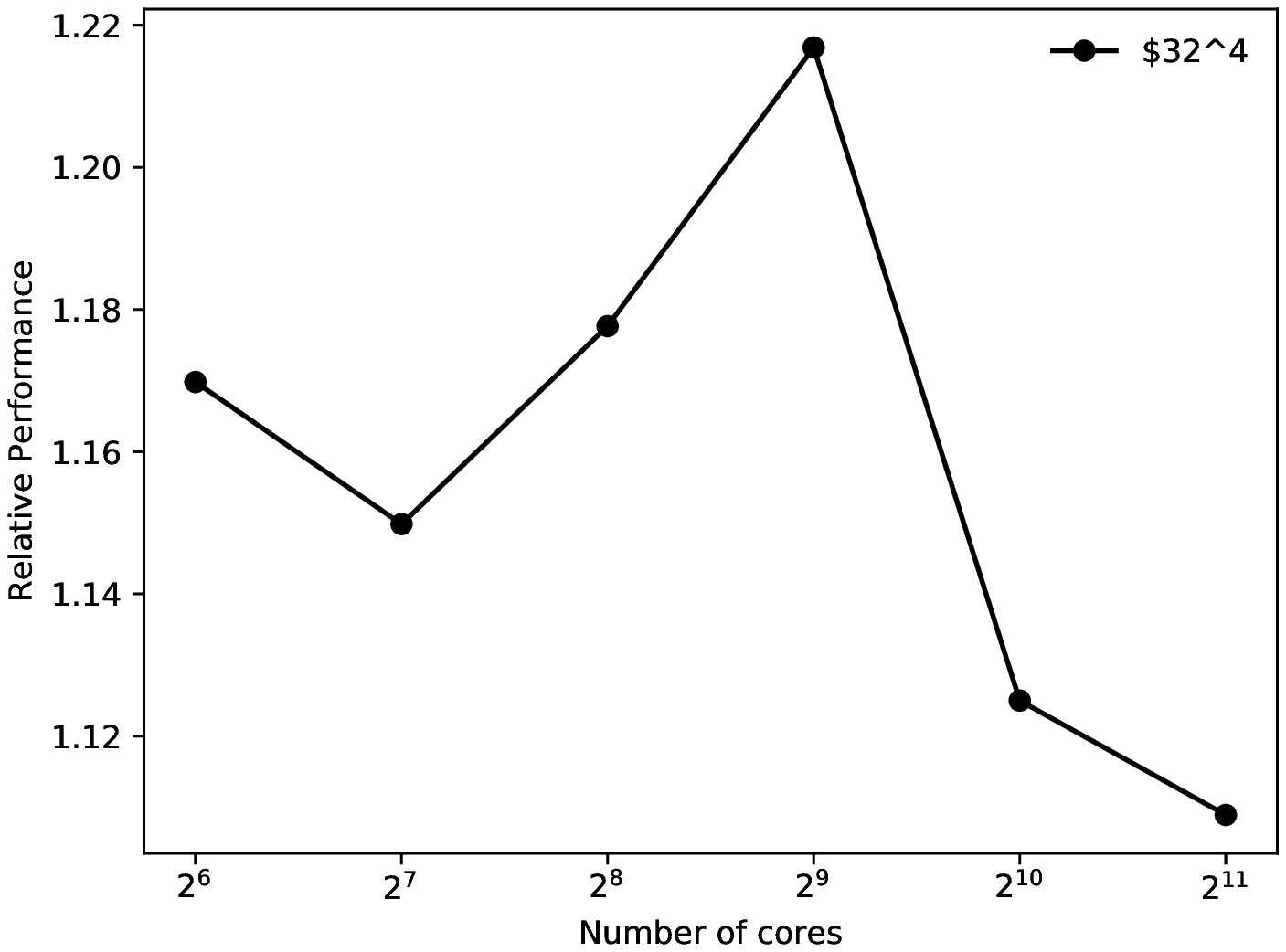}
 \caption{ Left: Strong scaling performance on the Sunbird Skylake cluster measured against the AVX2 implementation. Right: Strong scaling performance on the Marconi Knights Landing cluster measured against the AVX2 implementation  }
 \label{fig:strongscaling_skl}
\end{figure}

\begin{figure}
\centering
\includegraphics[width=0.49\linewidth]{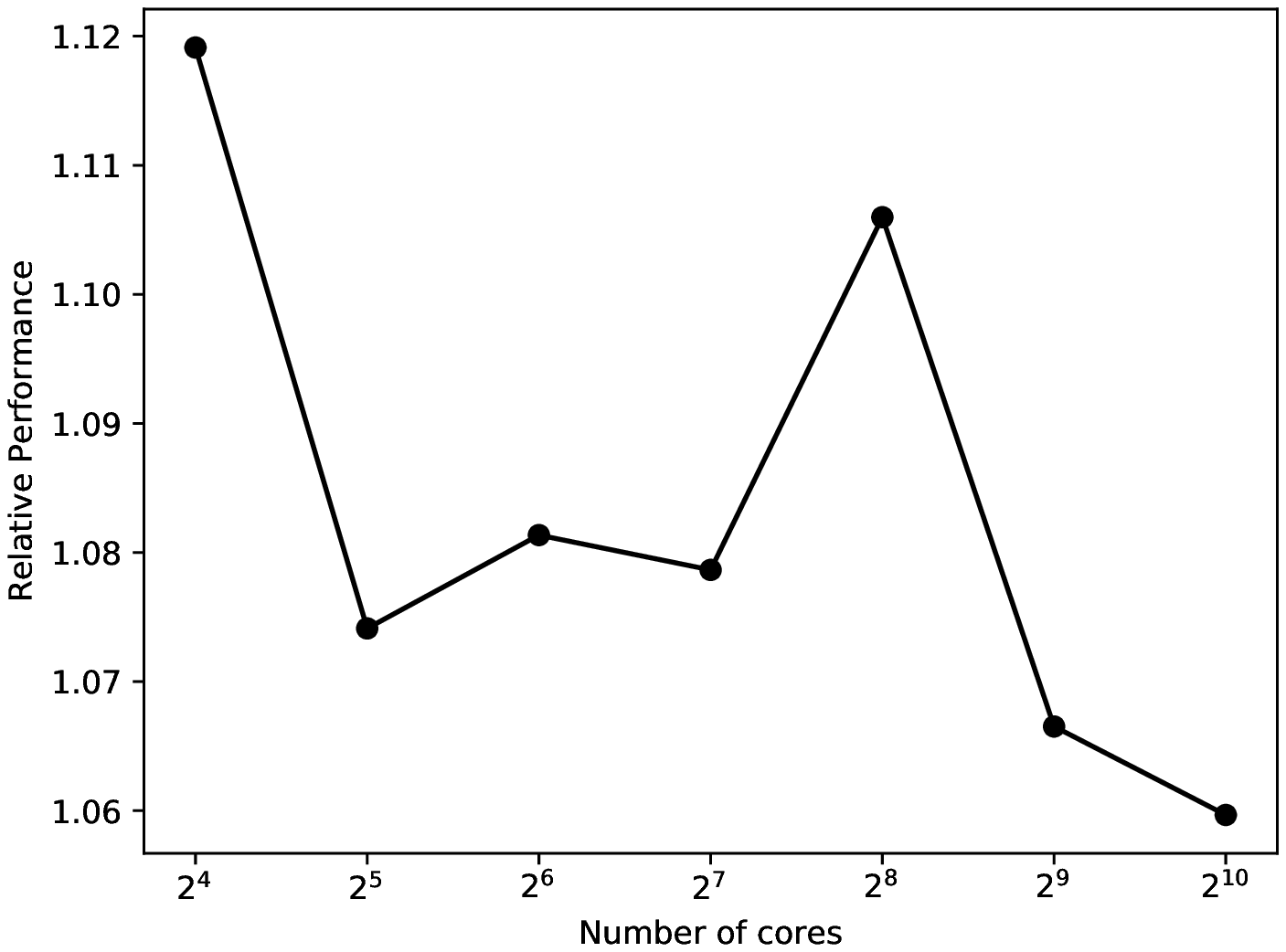}
\includegraphics[width=0.49\linewidth]{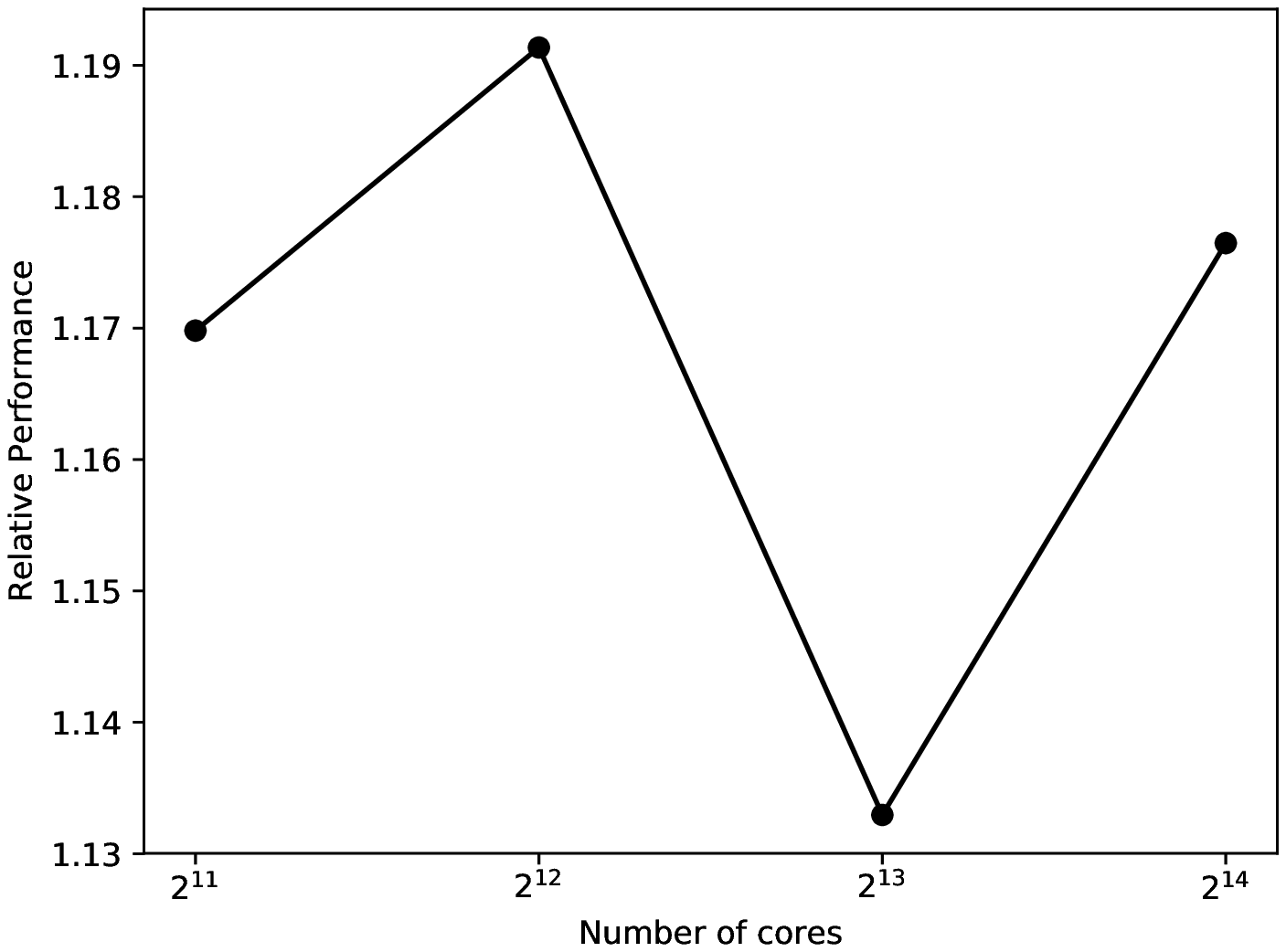}
\caption{ Left: Weak scaling performance on the Sunbird Skylake cluster using $n$ cores and the lattice size $V=48^3\times n$ measured against the AVX2 implementation.
Right: Weak scaling performance on the Marconi Knights Landing cluster using $n$ cores the lattice size  $V=32^3\times n$ measured against the AVX2 implementation. }
\label{fig:weakscaling}
\end{figure}

To get a more complete picture we measure the average time taken to generate a HMC trajectory. We have produced 6 trajectories starting from a random gauge configuration and report the average time per trajectory.
We use a Wilson-Yukawa gauge action with $\beta=1.5$, $c_0=5/3$ and $c_1=-1/12$ and include two fermions with $\kappa=0.278000465$ and $0.276509194$ and $c_{sw}=1$. Two levels of smearing are enabled in the fermion actions. Domain Deflation and blocking are enabled.

\begin{table} \centering
\begin{tabular}{|c|c|c|c|c|c|}
\hline
$ L $ & $ T $ & $N$ &  \multicolumn{2}{|c|}{ Time (s) / trajectory } & Speedup \\
 & & & AVX-512 & AVX2  & \\
\hline
32 & 32  & 1  & 1.59e3 & 1.86e3 & 1.17 \\
32 & 32  & 2  & 8.41e2 & 9.67e2 & 1.15 \\
32 & 32  & 4  & 4.39e2 & 5.17e2 & 1.18 \\
32 & 32  & 8  & 2.49e2 & 3.03e2 & 1.22 \\
32 & 32  & 16 & 1.52e2 & 1.71e2 & 1.13 \\
32 & 32  & 32 & 1.01e2 & 1.12e2 & 1.11 \\
\hline
32 & 32  & 1  & 1.59e3 & 1.86e3 & 1.17 \\
32 & 64  & 2  & 1.62e3 & 1.93e3 & 1.19 \\
32 & 128 & 4  & 1.73e3 & 1.96e3 & 1.13 \\
32 & 256 & 8  & 1.70e3 & 2.00e3 & 1.18 \\
\hline
\end{tabular}
\caption{ \label{knltimingtable} Average timings per trajectory using $N$ Knights Landing nodes with the volume $T\times L^3$. Speedup is measured against the AVX2 implementation. }
\end{table}

\begin{table} \centering
\begin{tabular}{|c|c|c|c|c|c|c|}
\hline
$ L $ & $ T $ & $N$ & $n$ &  \multicolumn{2}{|c|}{ Time (s) / trajectory } & Relative  \\
 & & & & AVX-512 & AVX2  & Speedup \\
\hline
32 & 48 & 1   &  16   & 1.31e3 & 1.43e3 & 1.09 \\
32 & 48 & 1   &  32   & 7.96e2 & 8.58e2 & 1.08 \\
32 & 48 & 2   &  64   & 3.96e2 & 4.28e2 & 1.08 \\
32 & 48 & 4   &  128  & 1.99e2 & 2.14e2 & 1.08 \\
32 & 48 & 7   &  256  & 1.16e2 & 1.23e2 & 1.06 \\
32 & 48 & 13  &  512  & 6.04e1 & 6.41e1 & 1.06 \\
32 & 48 & 26  &  1024 & 2.91e1 & 3.13e1 & 1.08 \\
\hline
32 & 24 & 1   &  8   &  2.48e2 & 2.79e2 & 1.13 \\
32 & 24 & 1   &  16  &  1.54e2 & 1.73e2 & 1.12 \\
32 & 24 & 1   &  32  &  9.55e1 & 1.03e2 & 1.08 \\
32 & 24 & 2   &  64  &  4.66e1 & 5.06e1 & 1.09 \\
32 & 24 & 4   &  128 &  2.36e1 & 2.59e1 & 1.10 \\
\hline
48 & 16   & 1   &  16   & 6.38e2 & 7.146e2& 1.12 \\
48 & 32   & 1   &  32   & 7.96e2 & 8.55e2 & 1.07 \\
48 & 64   & 2   &  64   & 7.99e2 & 8.64e2 & 1.08 \\
48 & 128  & 4   &  128  & 8.01e2 & 8.64e2 & 1.08 \\
48 & 256  & 7   &  256  & 8.87e2 & 9.81e2 & 1.11 \\
48 & 512  & 13  &  512  & 9.32e2 & 9.94e2 & 1.07 \\
48 & 1024 & 26  &  1024 & 9.72e2 & 1.03e3 & 1.06 \\
\hline
\end{tabular}
\caption{ \label{skltimingtable} Average timings per trajectory using $N$ Skylake nodes with $n$ cores and the volume $T\times L^3$. Speedup is measured relative to the AVX2 implementation.}
\end{table}

The timings for several lattice sizes and configurations of nodes are given in Tabs. \ref{knltimingtable} and \ref{skltimingtable} and shown in Fig. \ref{fig:strongscaling_skl} and \ref{fig:weakscaling}. Full compilation and runtime parameters and the simulation output are publicly available \cite{zenodo_skl_data} and \cite{zenodo_knl_data}.
The two version of the code scale similarly, with the AVX-512 version remaining faster in each case.   On the Sunbird Skylake machine, in a full trajectory the improvement is between 6\% and 13\%. Each node has 40 cores and a minimal number allocation of nodes is used in each case. On the Marconi KNL system the improvement is between 11\% and 22\%. In this case all 64 cores are used on each node. No clear dependence on the lattice size or number of nodes can be deduced from the data.

\begin{table} \centering
\begin{tabular}{|c|c|c|c|c|c|c|c|}
\hline
$ L $ & $ T $ & $N$ & $n$ &  \multicolumn{2}{|c|}{ Time (s) / trajectory } & speedup \\
 & & & & AVX-512 & AVX2  & \\
\hline
\multicolumn{7}{|l|}{ Skylake } \\
\hline
32 & 24 & 1 & 32  & 2.28e3 & 2.47e3 & 1.08 \\
32 & 24 & 2 & 64  & 1.12e3 & 1.21e3 & 1.08 \\
32 & 24 & 4 & 128 & 5.68e2 & 6.22e2 & 1.10 \\
\hline
\end{tabular}
\caption{ \label{gen2ltable} Average timings starting from a thermalised configuration per trajectory with the volume $T\times L^3$ with $N$ nodes using $n$ cores. }
\end{table}

Finally, we perform the same test with a thermalised starting configuration and a light quark. Two fermions are included $\kappa=0.27831$ and $0.276509$. The results are reported in Table \ref{gen2ltable}. The speedup achieved is similar to the previous tests, between 8\% and 10\%.

\section{Conclusion}

We announce an open source implementation of the Dirac operator in OpenQCD 1.6 with extended AVX-512 vector operations using Intel's intrinsic operations. These operations allow the application to make full use of the wider, 512-bit registers, reducing the total number of memory request, in particular those to L1 cache, and the number of floating point operations.

The implementation assumes that memory bandwidth is the main bottleneck in the application. Tradeoffs that reduce memory use at the cost of floating point operations and vector shuffles are considered acceptable. The application performs significantly better than the existing AVX2 implementation on Knights Landing and Skylake processors. In realistic benchmarking cases the improvement factor is between 6\% and 12\% on Skylake nodes and 11\% and 22\% on Knight Landing nodes.

\section{Acknowledgements}

We acknowledge the support of the Supercomputing Wales project, which is part-funded by the European Regional Development Fund (ERDF) via Welsh Government.

\end{document}